\documentclass[fleqn,twoside]{article}
\usepackage{espcrc2}

\usepackage{graphicx}
\usepackage[figuresright]{rotating}
\usepackage{amsfonts,amssymb}

\newcommand{\emiss}{$E_m$}
\newcommand{\pmiss}{$p_m$}

\title{First measurement of the spectral function at high energy and momentum in medium-heavy nuclei}

\author{D. Rohe\thanks{Daniela.Rohe@unibas.ch}\address[bs]
{University of Basel,  CH--4056 Basel, Switzerland} for the E97-006 collaboration}
       
\begin{document}

\begin{abstract}
The experiment E97-006 was performed at Jefferson Lab to measure the momentum and energy distribution of protons in the nucleus far from the region of the (approximate) validity of the mean field description, i.e. at high momentum and energies. The occurrence of this strength is long known from occupation numbers less than one.  In the experiment reported here this strength was directly measured for the first time. The results are compared to modern many-body theories. Further the transparency factor of $^{12}$C was determined in the $Q^2$-region of 0.6 to 1.8 (GeV/c)$^2$.  
\vspace{1pc}
\end{abstract}

\maketitle

\section{INTRODUCTION}
The energy and momentum distribution of nucleons bound in the nucleus are obviously determined by the nucleon-nucleon potential. The N-N potential consists of a repulsive core at small nucleon-nucleon distances of less than $\approx$ 1 fm and an attractive part. Such a realistic potential is usually constructed from N-N scattering data. In a vector meson exchange model the long-range part is represented by an exchange of a virtual pion whereas the short-range part is provided by heavier mesons like the omega and the rho. The strongly repulsive part, which is caused by the short-range and tensor components of the interaction, is ignored in the mean field description. In the Independent Particle Shell Model (IPSM) it is assumed that the nucleons move independently from each other. The repulsive part responsible for N-N correlations is neglected. Therefore all nucleons reside below the Fermi energy and Fermi momentum of about 250 MeV/c. All orbits are fully occupied in the IPSM. It was somewhat of a surprise when (e,e'p) measurements performed at NIKHEF \cite{Wit90} found values of typically 65~\% for the spectroscopic factor of the valence orbits in nuclei from lithium to lead, {\em ie.} 35~\% are missing. Short-range  (SRC) and tensor correlations are responsible for up to 15~\% depletion. Due to the repulsive core of the interaction at small distances nucleons are scattered to high energy and momentum. Another 20~\% is coming from long-range correlations (LRC) which lead to a fragmentation of the single-particle states due to coupling to collective modes like vibration. This influences the spectral function in particular at small energy, {\em i.e.} the valence states. For the deeper lying orbits the occupation is larger.

Modern many-body theories like the Correlated Basis Function theory (CBF) \cite{Ben89}, the Green's function approach (GF) \cite{Mue95} and the self-consistent Green's function theory (SCGF) \cite{Fri03} are able to deal with the realistic N-N potential directly and therefore contain short-range and tensor correlations. The remarkable difference to the momentum and energy distribution derived in the IPSM is the additional strength found at large energy E and momentum k. Above the Fermi momentum $k_F$ the single-particle strength is negligible. As a consequence the search for SRC has to concentrate on large E and k. A practical tool is the (e,e'p) experiment.

\section{EXPERIMENT}
\begin{figure}[t]
\begin{center}
\includegraphics[width=6cm,clip]{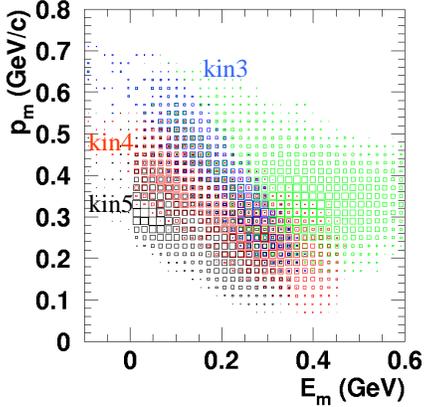}
\caption{\label{empm}Coverage of the $E_m$,$p_m$-plane by the runs taken in
parallel kinematics shown in a cross section times phase space plot.(Due to the large momentum acceptance of the spectrometers,
part of the data (green) are for $\theta_{kq}>45^\circ$).}
\end{center}
\end{figure}

 The experiment was performed in Hall C at Jefferson Lab employing three
quasi-parallel and two perpendicular kinematics at a q $\gtrsim$ 1~(GeV/c) 
(for a detailed discussion see \cite{Rohe04}). Electrons of 3.3~GeV energy 
 and beam currents up to 60~$\mu$A were incident upon $^{12}$C, $^{27}$Al, 
$^{56}$Fe and  $^{197}$Au targets.  The scattered electrons were detected in the HMS spectrometer (central momenta 2 - 2.8~GeV/c), the protons were detected in the SOS spectrometer (central momenta 0.8 - 1.7~GeV/c). Fig.~\ref{empm} gives the kinematical coverage for the parallel kinematics.  

From the measured scattering angles and momenta of the electron and the proton the missing energy 
\begin{equation}
E_m = E_e - E_{e'} - T_{p'} - T_{A-1}
\end{equation} 
and the missing momentum
\begin{equation}
\vec{p}_m = \vec{q} - \vec{p}_{p'}
\end{equation} 
can be constructed. Here $E_e$ is the electron beam energy, $E_{e'}$ the energy of the scattered electron and $T_{p'}$ ($T_{A-1}$) the kinetic energy of the knocked-out proton (the residual nucleus). In Plane Wave Impuls Approximation (PWIA) there is a direct relation between the measured quantities and the theory. In this case \emiss\ can be identified with the removal energy $E$ of the proton in the nucleus and \pmiss\ with its initial momentum $-\vec{k}$. However, in the nucleus the hit proton can interact with the other nucleons of the nucleus. Obviously this rescattering contribution is reduced for large momentum transfer q and increases for heavier nuclei. Calculations of the rescattering contribution \cite{Barb05} confirm that rescattering is small for $^{12}$C in parallel kinematics where the momentum transfer $\vec{q}$ is parallel to the initial momentum of the proton.  The data taken in perpendicular kinematics lead to a three times larger strength integrated over the experimental acceptance and compared to the
parallel kinematics. Rescattering contributions cannot entirely explain this  difference. Probably multi-step processes and Meson Exchange Currents (MEC) need to be considered. Work in this direction is underway \cite{Barb_nu,Barb}. 

The raw data were analyzed using two different procedures, both based on an
iterative approach and a model spectral function. In one, the phase
space is taken from a Monte Carlo simulation of the experiment, and the spectral
function is determined from the acceptance corrected cross sections. 
Radiative corrections are taken into account according to
\cite{Ent01}. The other is based on a bin-by-bin comparison of experimental and Monte Carlo yield, where the Monte Carlo simulates 
the known radiative processes, multiple scattering and energy loss of the
particles using spectrometer transfer matrices. The parameters of the model spectral
function  then are iterated to get agreement between data and simulation.  We have found good agreement between the two procedures.

In PWIA the spectral function can be obtained via the absolute differential cross section 
\begin{equation} \label{cross}
\frac{d\sigma}{d\Omega_{e}\,d\Omega_{p}\,dE_{e'}\,dE_{p'}} = K \, \sigma_{ep} \, S(E_m,\vec{p_m}) \, T_A(Q^2).
\end{equation} 
Here $K$ is a kinematical factor and $\sigma_{ep}$ the e-p cross section for a moving proton bound in the nucleus. Since the bound proton is off-shell, {\em i.e.} the usual relation between mass, energy and momentum is not longer valid, the expression for the e-p cross section cannot be derived uniquely. Different versions are on the market like $\sigma_{cc1}$ and $\sigma_{cc2}$ \cite{For83} which are derived from two expressions of the electromagnetic current operator $\Gamma_1$ and $\Gamma_2$. Both are equivalent in the on-shell case but differ for an off-shell proton. In the following the  e-p cross section $\sigma_{cc}$ \cite{Rohe04} will be used where the kinematical variables are not modified to obtain an equivalent on-shell kinematics as in the case of $\sigma_{cc1}$ and $\sigma_{cc2}$. The difference between the cross sections is small in the IPSM region but increases with increasing \pmiss\ and \emiss. The choice of $\sigma_{cc}$ is motivated by the fact that it gives the best agreement of the spectral function for same \emiss\ and \pmiss\ but measured in different (parallel) kinematics. 

The absorption of the proton in the nucleus is accounted for by the nuclear transparency $T_A$. Also for a check the analysis procedure $T_A$ was measured for five different kinematics on $^{12}$C. The results are presented in the next section and compared to the correlated Glauber theory of Benhar \cite{Rohe05}.

\section{TRANSPARENCY}
\begin{figure}[t] 
\begin{center}
\includegraphics[width=6.cm,clip]{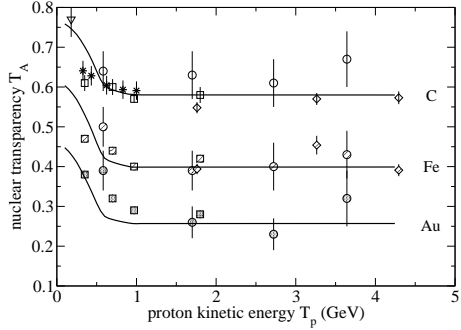}
\caption{\label{res_trans}Nuclear transparency $T_A$ for C, Fe and Au as a function of the proton kinetic energy $T_p$ compared to the correlated Glauber calculations (solid lines). The data indicated by circles are from the NE18--experiment at SLAC \cite{Nei95}, squares and diamonds are Jlab data of \cite{Abb98} and \cite{Gar02} and from Bates \cite{Garino92} (triangle down). The result indicated by stars is obtained with the correlated spectral function of \cite{Ben94}.}
\end{center}
\end{figure} 
Measurements of the transparency rely on comparison of the yield measured for a specific nucleus to the yield calculated assuming PWIA. This transparency accounts for absorption of the proton in the nucleus but also for large momentum changes due to an interaction with the surrounding nucleons. The  calculated yield $N^{sim}(E_m,p_m)$ is taken from a Monte Carlo simulation which contains the detector acceptances, radiative corrections etc. It is checked that the spectra from the simulation are in good agreement with the one measured or reconstructed from measured quantities. As input for the simulation a spectral function is needed. Because the spectral function is best known at small \emiss\ and \pmiss, one restricts this measurement to the IPSM region. As boundaries \emiss\ $<$ 80 MeV and \pmiss\ $<$ 300 MeV/c are chosen, the same as used in previous analyses at SLAC and Jlab \cite{Nei95,Abb98,Gar02}. Then the nuclear transparency is obtained by the ratio of the measured and simulated yield integrated over the chosen (\emiss,\pmiss) region:
\begin{equation} \label{extr_trans}
T_A(Q^2) =  \frac{\int_V d{\bf p}_m\, dE_m\, N^{exp}(E_m,{\bf p}_m)}{\int_V d{\bf p}_m\, dE_m\, N^{sim}(E_m,{\bf p}_m)}.
\end{equation} 
In the simulation a spectral function has to be chosen which should correctly describes the experimental distribution in the region of interest. In the previous analyses an IPSM spectral function was used which factorizes into an $E$ and $k$ distribution. The $E$-distribution is described by a Lorentzian and the momentum distribution is derived from a Wood-Saxon potential whose parameters are adjusted to data measured at Saclay \cite{Mou76}. This spectral function does not account for any depletion of the shells due to N-N correlations. To take this into account the simulated yield for carbon is divided by a factor of $\epsilon^{SRC}$ 1.11 $\pm$ 0.03. This factor was used in all the previous analyses and was estimated from an early examination of N-N correlations in $^{12}$C and $^{16}$O \cite{Sick93,Orden80}. The inverse of $\epsilon^{SRC}$ corresponds to the occupation number, which modern many-body theories predict to be lower, $\approx$ 80 - 85\% \cite{Ben89,Mue95}. This number is larger than the spectroscopic factor giving the occupation of a state; it contains the background strength, also caused by correlations, which cannot be attributed to a specific orbit \cite{Ben90,Mue04}. The above considerations emphasize that the number of nucleons missing due to absorption in the nucleus cannot be distinguished from the depletion of the single-particle orbits due to SRC. 

\begin{figure}[t] 
\begin{center}
\includegraphics[width=6.5cm,clip]{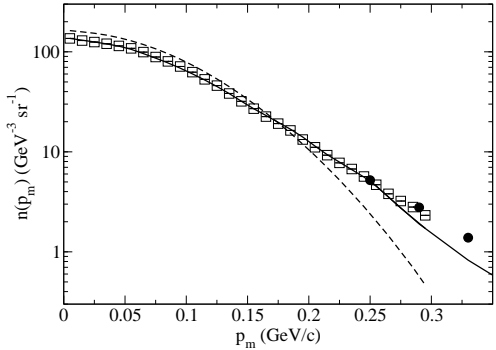}
\caption{\label{s_mom}Momentum distribution in the region of 0.03~GeV $<$ \emiss\ $<$ 0.08~GeV obtained from the data, the CBF theory (solid) and the IPSM (dashed). Three data points (circles) are from data focusing on the high \pmiss\ region \cite{Rohe04,Rohe04a} (s. sec. \ref{sec_spec}).}
\end{center}
\end{figure}  

The factor $\epsilon^{SRC}$ introduced in the IPSM spectral function to account for the depletion due to N-N correlations is certainly an approximation. Comparison to data for higher \pmiss\ (but still smaller than 300 MeV/c) reveal that the IPSM spectral function is still missing strength at \emiss\ around 0.05 GeV. To improve upon this approach a spectral function containing SRC from the beginning is employed in Eq.(\ref{extr_trans}). This spectral function is composed of a part due to SRC which accounts for 22\% of the total strength as calculated in Local Density Approximation (LDA), and the IPSM spectral function mentioned above but reduced by a factor (1 -- 0.22) to ensure normalization. This approach circumvents the application of the extra factor $\epsilon^{SRC}$. In addition, much better agreement between measured and simulated \emiss\ spectra is achieved. Using this spectral function the five data points for the nuclear transparency of $^{12}$C (solid symbols) are obtained which are shown as a function of the proton kinetic energy in Fig. \ref{res_trans}  together with previous results obtained at SLAC \cite{Nei95} (circles)  and Jlab \cite{Abb98,Gar02} (squares and diamonds). The error bars shown in the figure contain the statistical and systematic uncertainty but not the model-dependent error of 4.7~\%. This applies also to the data points of the previous works. Since the previous experiments were analyzed using the same assumption and ingredients the model-dependent error is the same for them, while it is somewhat lower in the case of using the CBF spectral function.

The solid lines drawn in Fig. \ref{res_trans} are the result of the correlated Glauber theory \cite{Rohe05} which takes Pauli Blocking, dispersion effects and N-N correlations into account. For comparison also results from previous experiments \cite{Gar02,Nei95,Abb98} for iron and gold are shown. For all three nuclei and large proton kinetic energy ($>$ 1.5~GeV) the theory describes the data well within the error bars. At low energy there is remarkable agreement between theory and the experimental results obtained using the CBF spectral function.

In fig. \ref{s_mom} the momentum distribution, mainly covering the S$_{1/2}$ state in $^{12}$C (\emiss = 0.03 - 0.08 GeV) is shown for the data (squares), the CBF theory and the IPSM. Obviously the IPSM momentum distribution starts to fail above \pmiss\ $\approx$ 200 MeV/c. It exceeds the data for small \pmiss\ by about 20~\% because no correction for SRC is applied. The CBF theory combined with the spectral function from IPSM as described above is in good agreement with the data. No extra  renormalization factor is needed. 

\section{RESULTS AT HIGH ${\bf E_m}$ AND ${\bf p_m}$} \label{sec_spec}
\begin{figure}[t]
\begin{center}
\center\includegraphics[scale=0.25,clip]{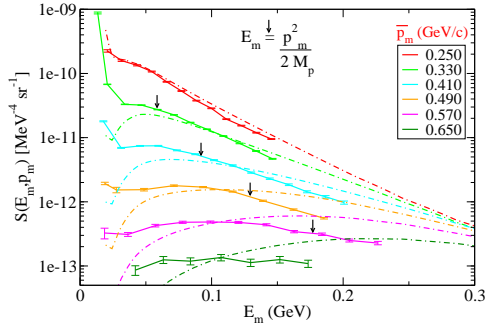}
\end{center}
\caption{\label{spec_frick}Spectral function  
for the $^{12}{\mathrm{C}}$ nucleus. Experimental result for several momenta above the Fermi
momentum are shown as solid lines with error bars. 
The dashed-dotted lines represent the SCGF nuclear matter spectral 
function at a density of $\rho=0.08\,\mbox{fm}^{-3}$ \cite{Fri03,Frick04}. }
\end{figure} 

\begin{figure}[t]
\begin{center}
\center\includegraphics[scale=0.25,clip]{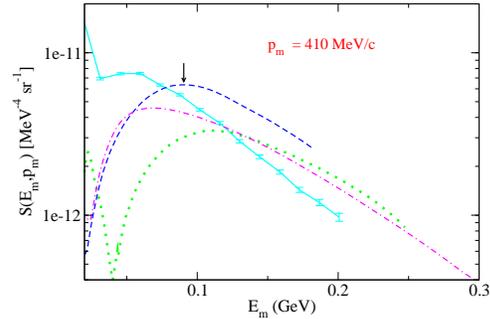}
\end{center}
\caption{\label{spec_comp} A comparison between the experimental result at 
$k=410\,\mbox{MeV/c}$ (solid lines with error bars), the
theoretical spectral functions  
by Benhar {\it et.~al.}~\protect{\cite{ben94}} (dashed line), the SCGF result (dashed-dotted line) and the second order Greens function approach \cite{Mue95} (dotted).}
\end{figure} 
To obtain the spectral function from eq. \ref{cross} a transparency factor of 0.6 was used for $^{12}$C. The results presented here are restricted to the data measured in parallel kinematics to keep the corrections on the distorted spectral function small. The region of the onset of the $\Delta$ resonance is clearly visible in the data and its contribution is reduced by a cut. The experimental spectral function measured for $^{12}$C is shown in fig. \ref{spec_frick} together with the theoretical result obtained in the SCGF approach for nuclear matter at a density comparable to $^{12}$C \cite{Fri03}. At low momentum, around the Fermi momentum, good agreement is observed.  Interestingly this is also the case for the spectral functions of refs. \cite{ben94,Mue95}. At higher \pmiss, however, there are significant differences as can be seen in fig. \ref{spec_comp}. The spectral function from CBF theory has its maximum at around $p_{m}^{2}/(2M)$ which would be the case for two nucleons leaving the nucleus after a hard collision due to SRC. However, the data show more strength at lower \emiss\ and the maximum of the spectral function is shifted to smaller \emiss\ values. This might be due to LRC in finite nuclei which are not treated in the nuclear matter calculation. Further tensor correlations which are known to be more important than central short range correlations might lead to a relocation of strength. Comparison of the two Green's function approaches shows a remarkable improvement of the self-consistent treatment compared to the one which contains only 2nd order diagrams. This indicate that higher order diagrams are important and more complicated configurations than a simple hard interaction between two nucleons are involved.

\begin{figure}[t]
\begin{center}
\includegraphics[scale=0.42]{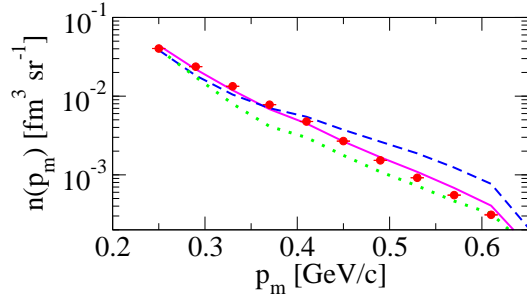}
\caption{\label{mom_comp}Momentum distribution of the data (circles) compared to the theory of refs. \cite{Mue95} (dots), \cite{Fri03} (solid) and \cite{ben94} (dashed). The lower integration limit is chosen as 40 MeV, the upper one to exclude the $\Delta$ resonance.}
\end{center}
\end{figure} 

From the spectral function the momentum distribution and the number of protons found in the (\emiss,\pmiss) region covered by the experiment can be obtained by integration. The lower integration limit is set to \emiss\  = 40 MeV to avoid contributions from the single-particle region and LRC contributions; the upper limit is adjusted in such a way to exclude the $\Delta$ resonance. The same cuts are employed for theory. In fig. \ref{mom_comp} the experimental momentum distribution is compared to the three theories. The result from the SCGF approach agrees best with the data whereas the other two calculations slightly over- or undershoot the data. However, one should not forget that due to the choice of the integration limits in \emiss\ the comparison depends somewhat on the different shapes of the experimental and theoretical spectral functions.

Integrating the spectral function over the (\emiss,\pmiss) region covered by the experiment (\pmiss\ = 0.23 - 0.67 GeV/c) and using the integration limits discussed above, one obtains the correlated strength. This strength represents the number of protons which are located outside the IPSM region. The numbers obtained from the data and from three theories are quoted in table \ref{strength}. To account for FSI a correction of --4~\% has to applied to the experimental value in table \ref{strength} \cite{Barb_nu,Barb04}. Good agreement is found between data and the CBF theory as well as with the SCGF approach. 

Comparison of the results to the ones obtained for heavier nuclei (Al, Fe, Au) shows that the shape of the spectral function for C, Al, and Fe ist quite similar. For Au a larger contribution from the broader resonance region is obvious and the maximum of the spectral function is shifted to higher \emiss. The correlated strength for Al, Fe and Au is 1.05, 1.12 and 1.7 times the strength for C normalized to the same number of protons. This increase cannot be solely explained by rescattering but MEC's have probably taken into account. Another contribution may be coming from the stronger tensor correlations in asymmetric nuclear matter \cite{Kon05,Has05}.   

\begin{table}[t!]
\begin{center}
\begin{tabular}{l|l}
\hline
\rule{0mm}{5mm} Experiment & ~~0.61 $\pm 0.06$ \\
~Greens function  theory \cite{Mue95}~~ & ~~0.46 \\
~CBF theory \cite{Ben89} & ~~0.64 \\
~SCGF theory \cite{Fri03} & ~~0.61\\[2mm] 
\hline 
\end{tabular}
\end{center}
\caption{\label{strength} Correlated strength (quoted in
terms of the number of protons in $^{12}$C.)}
\end{table}

\end{document}